\documentclass[twocolumn,english]{IEEEtran}
\usepackage[T1]{fontenc}
\usepackage{xcolor}
\usepackage{pdfcolmk}
\usepackage{babel}
\usepackage{float}
\usepackage{graphicx}
\PassOptionsToPackage{normalem}{ulem}
\usepackage{ulem}
\usepackage{subscript}
\usepackage[unicode=true,
 bookmarks=true,bookmarksnumbered=true,bookmarksopen=true,bookmarksopenlevel=1,
 breaklinks=false,pdfborder={0 0 0},pdfborderstyle={},backref=false,colorlinks=false]
 {hyperref}
\hypersetup{pdftitle={Your Title},
 pdfauthor={Your Name},
 pdfpagelayout=OneColumn, pdfnewwindow=true, pdfstartview=XYZ, plainpages=false}

\makeatletter

\providecommand{\tabularnewline}{\\}
\providecolor{lyxadded}{rgb}{0,0,1}
\providecolor{lyxdeleted}{rgb}{1,0,0}

\ifCLASSOPTIONcompsoc
\usepackage[caption=false,font=normalsize,labelfont=sf,textfont=sf]{subfig}
\else
\usepackage[caption=false,font=footnotesize]{subfig}
\fi

\makeatother

\begin{document}

\title{Neutron-induced strike: Study of multiple node charge collection
in 14nm FinFETs}

\author{Nanditha P. Rao and Madhav P. Desai\\
Indian Institute of Technology Bombay\\
Email: \{nanditha@ee, madhav@ee\}.iitb.ac.in}
\maketitle
\begin{abstract}
FinFETs have replaced the conventional bulk CMOS transistors in the
sub-20nm technology. One of the key issues to consider is, the vulnerability
of FinFET based circuits to multiple node charge collection due to
neutron-induced strikes. In this paper, we perform a device simulation
based characterization study on representative layouts of 14nm bulk
FinFETs in order to study the extent to which multiple transistors
are affected. We find that multiple transistors do get affected and
the impact can last up to five transistors away (\textasciitilde{}200nm).
We show that the potential of source/drain regions in the neighborhood
of the strike is a significant contributing factor. In the case of
multi-fin FinFETs, the charge collected per fin is seen to reduce
as the number of fins increase. Thus, smaller FinFETs are susceptible
to high amounts of charge collection.
\end{abstract}

\begin{IEEEkeywords}
multiple transients, layout approach, critical area, soft error
\end{IEEEkeywords}

\section{\label{sec:Introduction}Introduction}

\IEEEPARstart{ W}{hen} high energy particles such as alpha particles
or neutrons strike a semiconductor device, they generate charge at
the region of strike. This charge can either recombine or get collected
in the source/drain regions through diffusion or other mechanisms
\cite{SEU_physics1,combn1_glitch_ht,SEU_phy4}, resulting in a transient
current known as single event transient (SET). A particle strike on
a device layout can generate SETs in multiple transistors across different
standard cells, affecting multiple logic gates. These SETs can propagate
to flip-flops and flip the stored value resulting in multiple bit-flip
errors, known as soft errors. The rate at which these errors occur
is known as soft error rate (SER).

Typically, the reliability estimates at the architectural level are
calculated based on a single-bit-flip fault model for soft errors
in which a single random bit is expected to flip at any time \cite{single_bit1,single_bit_flip_element,single_flip_arch}.
Similarly, at the circuit level, an SET is modeled as a current injection
into the drain of a single transistor \cite{double_exp_drain_cur1,drain_cur2}.
If multiple transients and multiple flips were to occur, these models
will not accurately represent the reality and will result in optimistic
reliability estimates. Therefore, it is important to quantify the
extent to which a circuit/layout is susceptible to multiple transients
especially in the current technology.

Device layouts with FinFETs, which have replaced the planar MOSFETs
for sub-22nm technologies, are also susceptible to SETs. Most existing
studies on soft errors in FinFETs focus on memories and show that
the radiation sensitivity of FinFET based SRAMs is better than that
of planar SRAMs \cite{finfet_SRAM_drain,FinFet_planar1_drain}. This
is mainly attributed to the fact that the volume of the source/drain
region (the fin) that connects to the substrate is small as compared
to planar devices, resulting in reduced charge collection. With technology
scaling, a radiation-induced strike can have a large region of influence
and can affect multiple transistors and logic gates. The phenomenon
of a radiation-induced strike affecting multiple transistors has been
studied to some extent in \cite{90nm_charge_sharing,mult_sharing2,mult_130nm_sec_energy}
for planar MOSFETs but has not been understood to the same extent
in FinFETs. Some studies with respect to FinFETs are performed in
\cite{finfet_multi1,finfet_multi2} and they report that multiple
cell upsets do occur in FinFET based SRAMs. Further, bulk FinFET based
designs are reported to have higher soft error rate than that of SOI
based designs \cite{finfet_SOI_bulkSRAM}. Studies in \cite{FinFet_planar1_drain,Layout-placement,finfet_laser_ion_drain}
insist that better understanding of layout effects is necessary to
predict multiple event transients. In this paper, we confirm the already
known fact that multiple transients do occur, and we add value by
quantifying the fraction of a layout that is affected due to a single
particle strike. 

We find that a strike can have an impact up to five transistors away
(nearly 200nm) from the strike location. A source/drain region which
is at a higher potential, collects higher amounts of charge. However,
the nearest two transistors are the ones that are most affected. In
the case of multi-fin FinFETs, charge collected per fin reduces as
the number of fins increase. Thus, FinFETs with smaller widths are
susceptible to high amounts of charge collection. Since the region
of influence is large, the problem cannot be entirely tackled using
simple layout techniques. A careful circuit-aware placement of small
vulnerable gates may be necessary. 

The rest of this paper is organized as follows. In Section \ref{sec:Analysis-on-a},
we describe the device construction. We explain the role of potentials
on charge collection in Section \ref{sec:The-process-flow}. In Section
\ref{sec:The-role-of}. We present the quantification of multiple
node charge collection in a layout of single-fin and multi-fin FinFETs
in Section \ref{sec:Multiple-node-charge}. We summarize and conclude
the paper in Section \ref{sec:Conclusion}.

\section{\label{sec:Analysis-on-a}Device characterization}

In the 2D and 3D analysis of planar transistors, we saw that the range
of impact of a particle strike is large and hence multiple transistors
are affected. Simple layout techniques did not help reduce the charge
collection in multiple transistors. To study the extent to which multiple
transistors are affected in an emerging technology, we perform particle
strike simulations of 14nm bulk FinFET devices in 3D. We study the
range of impact of a particle strike and the role of potentials on
charge collection.

\begin{figure}[!tbh]
\includegraphics[scale=0.35]{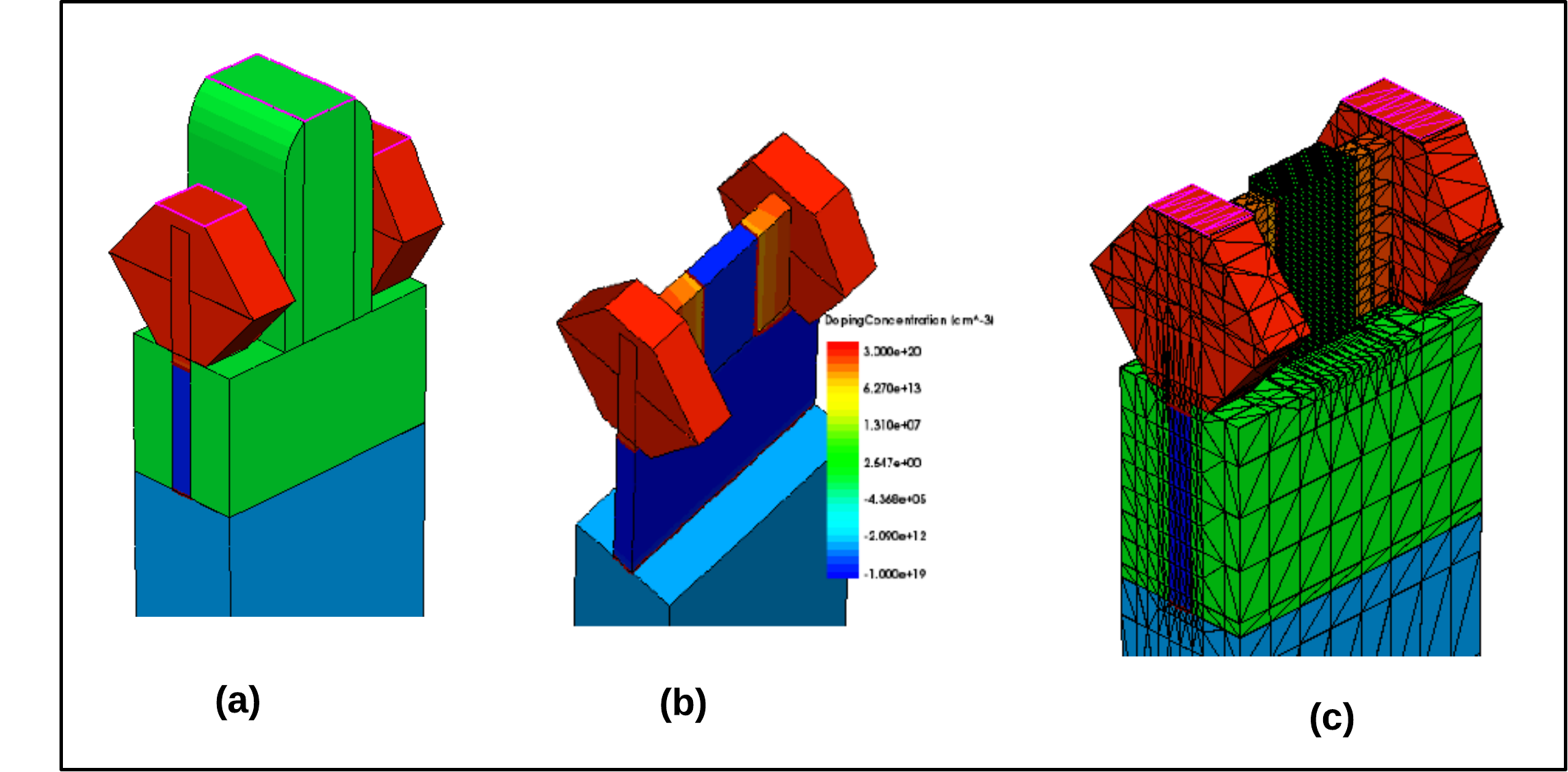}

\caption{\label{fig:finfet6}(a) Structure of the 14nm bulk n-FinFET. (b) The
fin is shown explicitly. (c) The FinFET device is shown along with
the mesh}
\end{figure}

In Figure \ref{fig:finfet6}, we show a bulk n-FinFET built based
on the information available in \cite{finfet_dev1,finfet_spec1}.
The geometry of the device and doping concentrations are shown in
Table \ref{tab:Geometry-and-doping}. The gate stack is constructed
with TiN/HfO2 \cite{finfet_WF_TiN}. Parameters such as the fin height,
STI depth, source/drain doping, channel doping and fin doping are
calibrated to meet the on/off current (I\textsubscript{on} and I\textsubscript{off})
and sub-threshold slope requirements of the 14nm bulk FinFET data
\cite{finfet_intel}. The upper fin has a gaussian doping profile
towards the channel as shown in the figure. The stress in the channel
region is modeled as a gaussian profile as per the data available
in \cite{finfet_stress1,finfet_stress2}. The device had a total of
78k elements after meshing.

\begin{figure}[!tbh]
$\;\;\;\;\;\;\;\;\;\;$\includegraphics[scale=0.4]{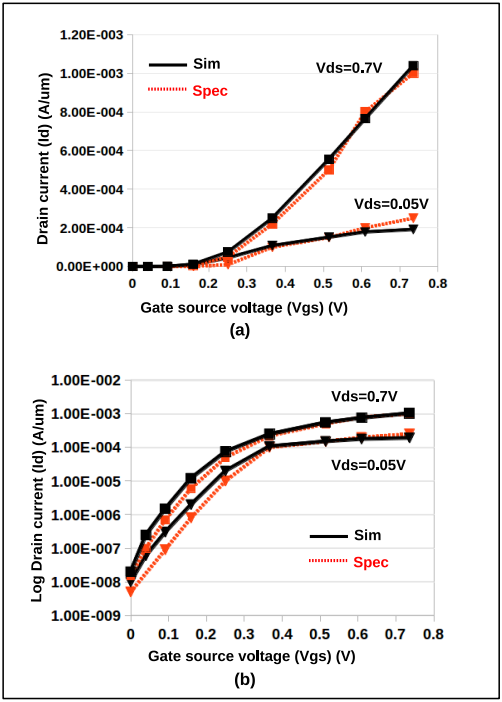}

\caption{\label{fig:IdVg_finfet}(a) Drain current(A/um) versus Gate source
voltage (Vgs) is plotted for a drain source voltage (Vds) of 0.7V
and 0.05V. The plot from our device is denoted by `Sim' and the one
in the specification \cite{finfet_intel} is denoted by `Spec'. (b)
log of Drain current(A/um) is plotted versus Gate source voltage (Vgs).}
\end{figure}

\begin{figure}[!tbh]
$\;\;\;\;\;\;\;\;\;\;\;\;\;\;$\includegraphics[scale=0.4]{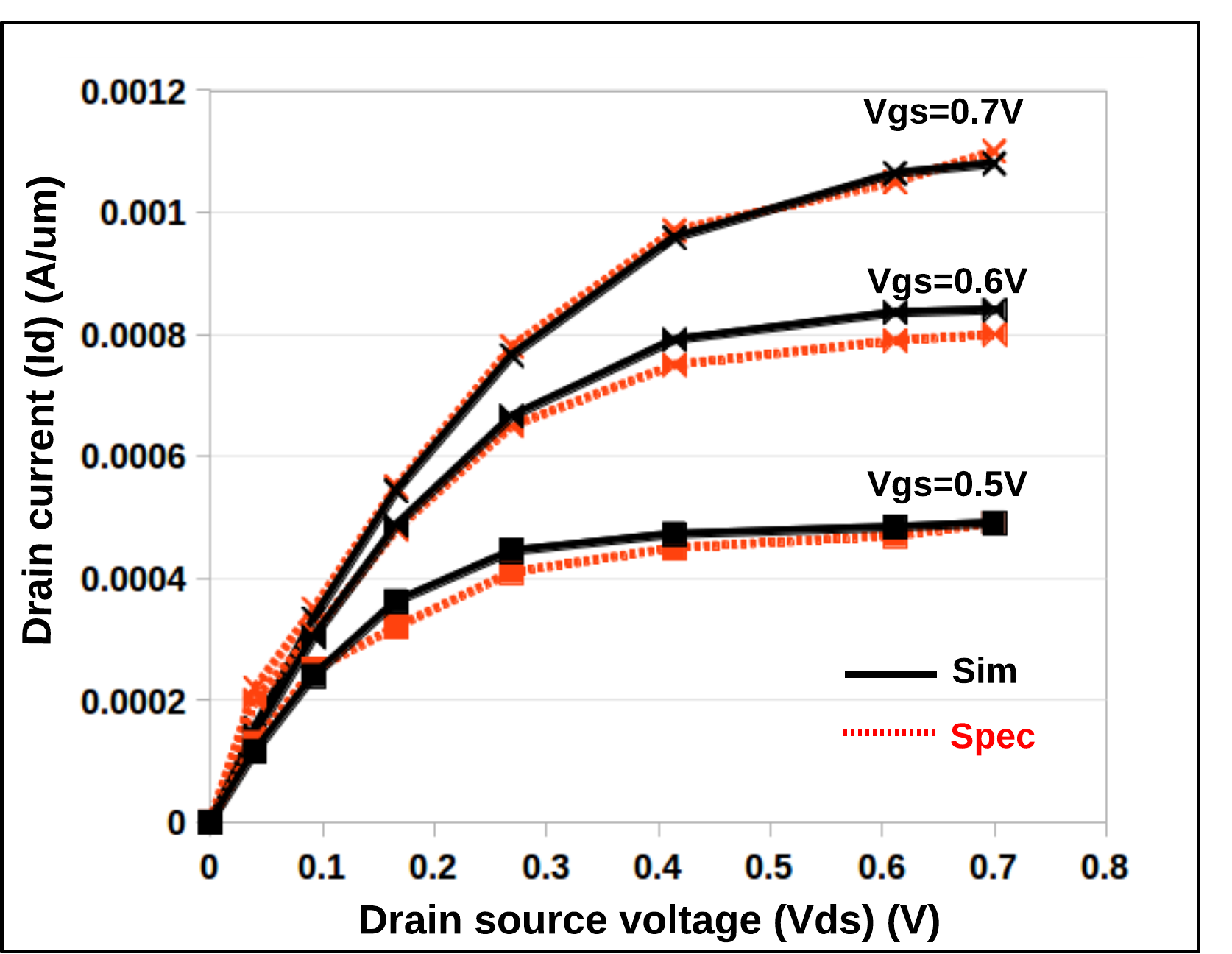}

\caption{\label{fig:finfet_IdVd} Drain current(A/um) versus Drain source voltage
(Vds) is plotted for gate source voltages (Vgs) of 0.5V, 0.6V and
0.7V. The plot from our device is denoted by `Sim' and the one in
the specification \cite{finfet_intel} is denoted by `Spec' .}
\end{figure}

\begin{table}[!tbh]
\begin{tabular}{|c|c|}
\hline 
Device parameter & Value\tabularnewline
\hline 
\hline 
Gate length & 20nm \tabularnewline
\hline 
Effective oxide thickness & 1.2nm \tabularnewline
\hline 
Fin height & 45nm\tabularnewline
\hline 
Fin width & 10nm\tabularnewline
\hline 
Fin pitch & 42nm\tabularnewline
\hline 
Gate pitch & 70nm\tabularnewline
\hline 
STI depth & 60nm\tabularnewline
\hline 
Total depth of the substrate & 400nm\tabularnewline
\hline 
Lower fin doping (Boron) & 1e19 cm\textsuperscript{-3}\tabularnewline
\hline 
Channel doping (Boron)  & 3e18 cm\textsuperscript{-3}\cite{finfet_channeldoping}\tabularnewline
\hline 
Source/drain doping (Arsenic) & 1e20 cm\textsuperscript{-3}\tabularnewline
\hline 
\end{tabular}

\caption{\label{tab:Geometry-and-doping}Geometry and doping concentrations
of the 14nm bulk n-FinFET}
\end{table}

In\textcolor{red}{{} }\textcolor{black}{Figure \ref{fig:IdVg_finfet},}
we show a plot of the drain current versus gate to source voltage
(V\textsubscript{gs}) of our device (denoted by `Sim') for a drain-source
voltage (V\textsubscript{ds}) of 0.7V and 0.05V, as compared with
that in the specification (`Spec') \cite{finfet_intel}. In\textcolor{red}{{}
}\textcolor{black}{Figure \ref{fig:finfet_IdVd}} we show a plot of
the drain current versus drain to source voltage (V\textsubscript{ds})
of our device (denoted by `Sim') for gate-source voltage (V\textsubscript{gs})
of 0.5V, 0.6V and 0.7V, as compared with that in the specification
(`Spec') \cite{finfet_intel}. 

We model the particle strike as a cylindrical column of charge with
a gaussian radial track and is simulated using the HeavyIon module
in \textit{Sentaurus Device}. The physics models used in the simulation
are as follows. Mobility degradation effects due to impurity scattering,
carrier-carrier scattering, high electric fields and mobility degradation
at the silicon-insulator interface are specified using the models:
PhuMob, CarrierCarrierScattering, HighFieldsaturation, inversion and
accumulation layer model (IALMob) and Enormal (Lombardi) respectively.
Generation and recombination processes of electron-hole pairs are
modeled using Auger, Band2Band and SRH recombination models. Quantum
effects at the semiconductor\textendash insulator interface are modeled
using the eMultiValley modified local-density approximation (MLDA)
model. These physics models are consistent with the models used in
\cite{sajesh_finfet}.

\section{\label{sec:The-role-of}The role of potentials on multiple node charge
collection}

\begin{figure}[H]
$\;\;\;\;\;\;\;\;\;\;$\includegraphics[scale=0.45]{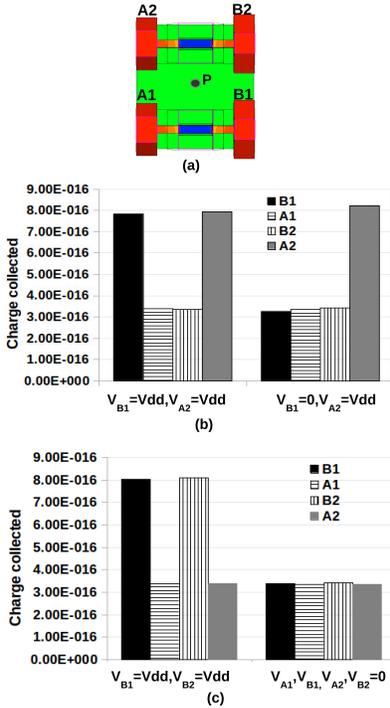}

\caption{\label{fig:Impact-of-voltagefin}(a) Layout of two bulk n-FinFETs
showing the particle strike location `P'. The source/drain regions
are marked as A1, A2, B1 and B2. (b) Charge collected in B1 and A2
is high when the respective nodes are biased high. (c) Charge collected
in B1 and B2 is high when the respective nodes are high. The least
charge collection is also shown.}
\end{figure}

To study the role of potentials of source/drain regions on charge
collection in the case of FinFETs, we setup the following experiment.
We construct a layout of two bulk n-FinFETs with a 11MeV particle
strike in between the two devices, as shown in Figure \ref{fig:Impact-of-voltagefin}
(a). We vary the voltages of all the source/drain regions and perform
particle strike simulations for each voltage combination. The charge
collected in each source/drain region (A1, B1, A2 or B2) is plotted
against the voltages in Figure \ref{fig:Impact-of-voltagefin} (b)
and (c).

We notice that the charge collected in a source/drain (A1, B1, A2
or B2) is high, when the terminal is biased high, and it reduces by
nearly 50\% when the terminal is biased low. For example, from Figure
\ref{fig:Impact-of-voltagefin} (b), we see that when B1 is biased
low, the charge collected in that node reduces by over 50\%. When
all nodes are biased low (Figure \ref{fig:Impact-of-voltagefin} (c)),
the charge collected in all the source/drain regions is minimum. This
observation is similar to that observed with the planar transistors.

\section{\label{sec:Multiple-node-charge}Multiple node charge collection
in a layout of 14nm bulk FinFETs}

\subsection{Charge collection in a layout of n and p FinFETs}

\begin{figure}[!tbh]
\includegraphics[scale=0.43]{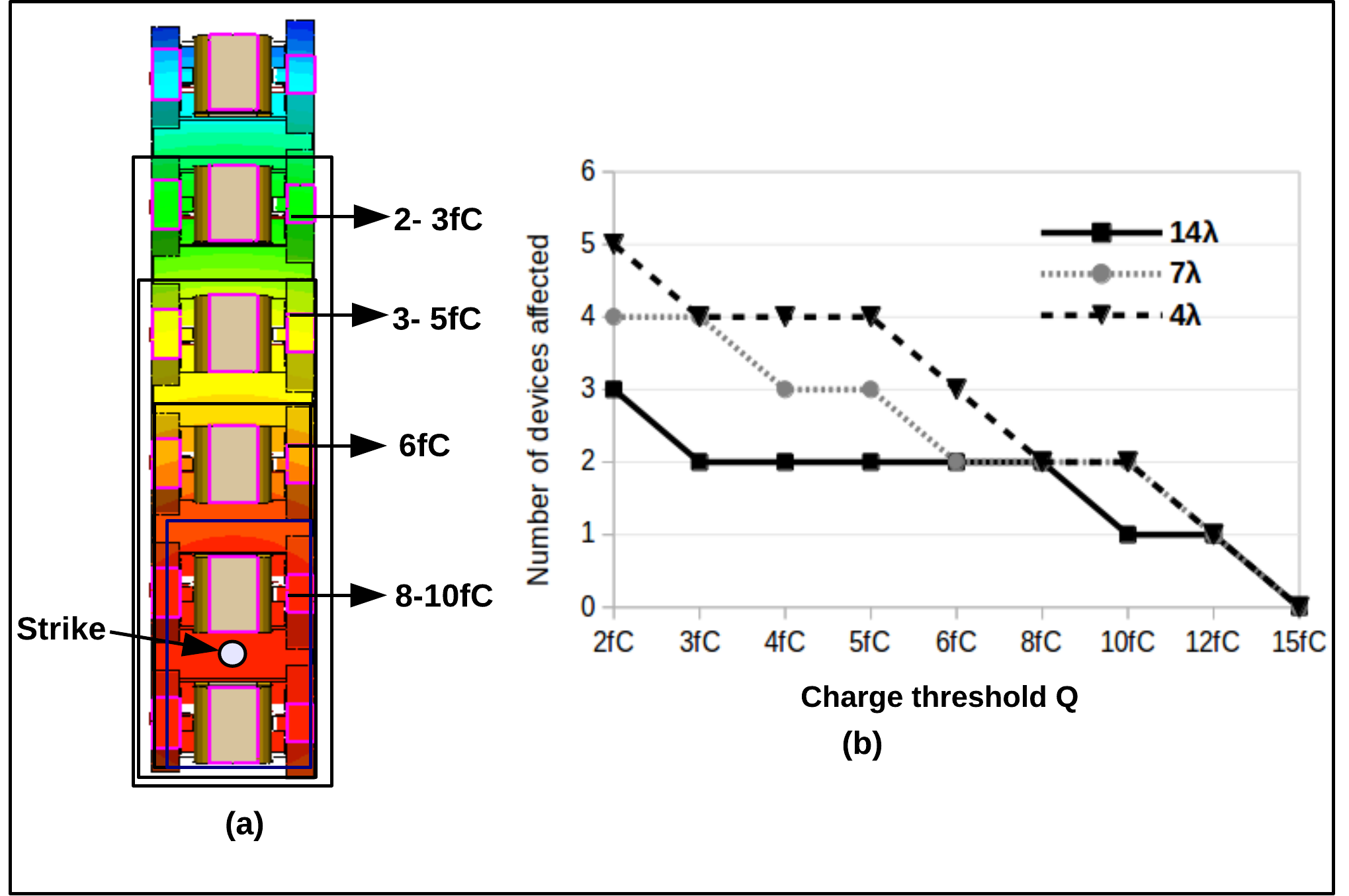}

\caption{\label{fig:Extent-to-which}(a) Layout of 6 bulk n-FinFETs showing
the strike location and charge collection map for a 11MeV particle
strike. (b) Graph showing the number of devices collecting more than
a certain charge Q (x-axis). }
\end{figure}

\begin{figure}[!tbh]
\includegraphics[scale=0.41]{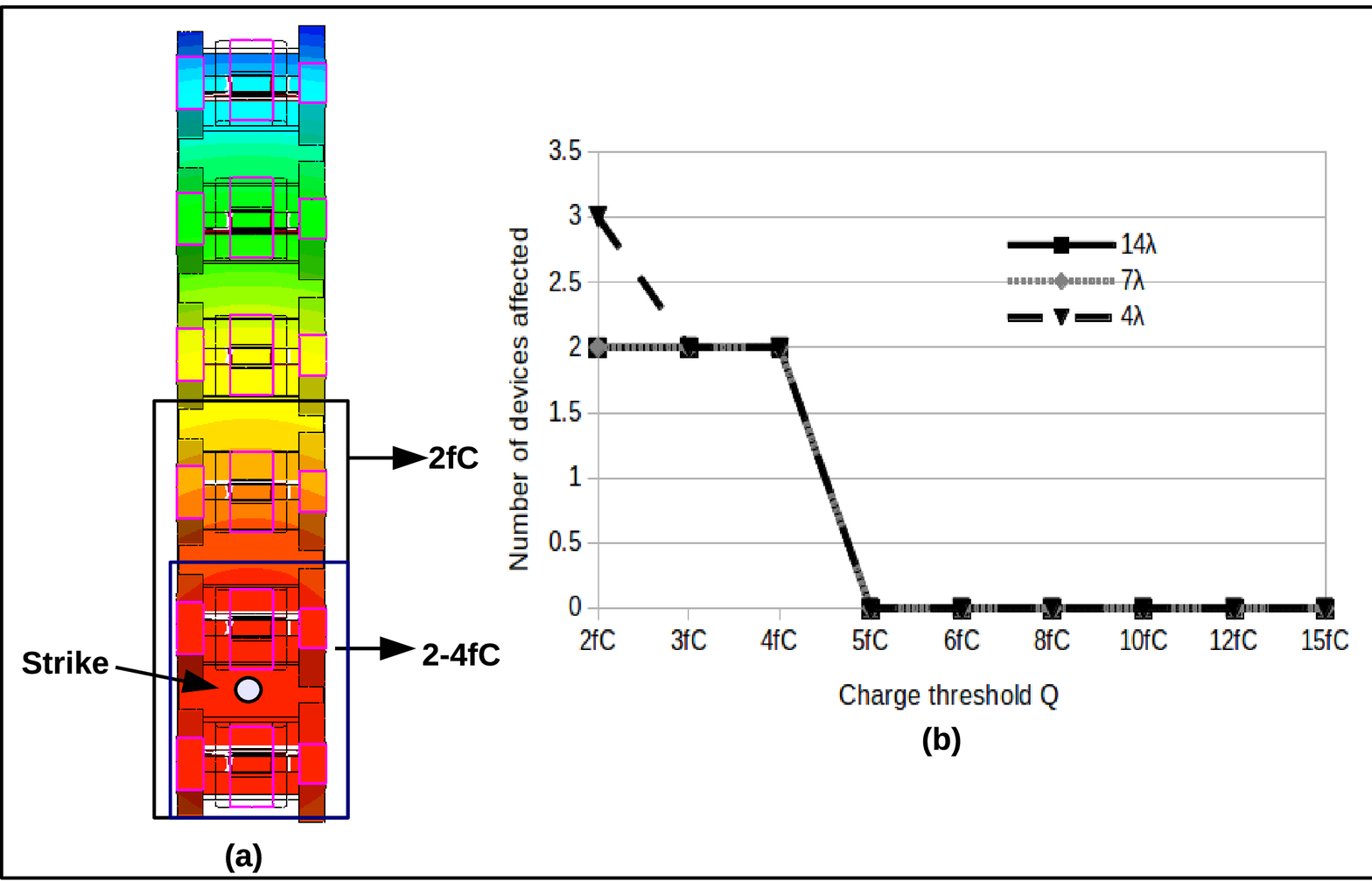}

\caption{\label{fig:Extent-to-which-1}(a) Layout of 6 bulk n-FinFETs showing
the strike location and charge collection map for a 5MeV particle
strike. (b) Graph showing the number of devices collecting more than
a certain charge Q (x-axis). }
\end{figure}

We construct a layout of six bulk n-FinFETs as shown in Figure \ref{fig:Extent-to-which}(a),
to study the extent to which a particle strike can affect multiple
transistors. Layouts with three device separations are simulated:
$4\lambda$, $7\lambda$ and $14\lambda$ (where $2\lambda=14nm)$.
We assume a particle to be incident in between the first two devices
as shown in the figure and measure the charge collected in the drain
regions (biased high to measure the maximum charge) of all the devices.
In Figure \ref{fig:Extent-to-which} and Figure \ref{fig:Extent-to-which-1},
we plot the collected charge against the number of devices which collect
at least so much amount of charge, for a particle energy of 11MeV
and 5MeV respectively. For a 11MeV strike, we see that up to five
devices (collect at least 2fC) can be affected due to a single particle
strike. The overall part of the layout that collects 2fC is marked
in the figure and the range of impact is nearly 200nm. However, the
nearest two devices collect maximum amount of charge (8-10fC). In
the case of 5MeV strike, two devices get affected on an average. Thus,
even in the case of FinFETs, a single particle strike can affect multiple
transistors and the region of the layout which is affected by the
strike is substantial. However, the nearest two devices are the ones
that are the most affected.

\begin{figure}[!tbh]
\includegraphics[scale=0.42]{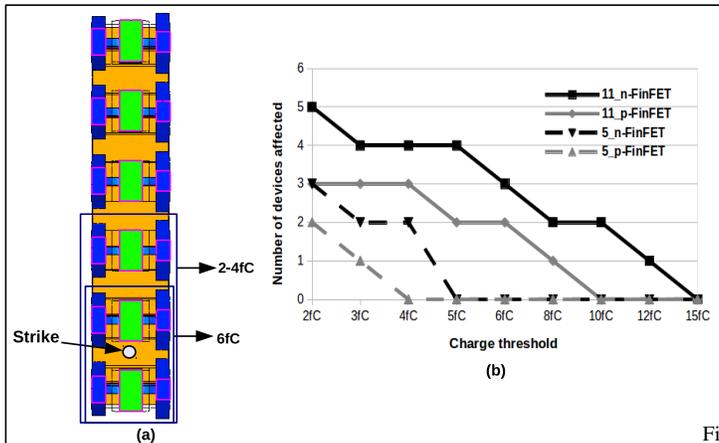}

\caption{\label{fig:p-finfets}\textcolor{red}{{} }(a) Layout of 6 bulk p-FinFETs
showing the strike location and charge collection map for a 11MeV
particle strike. (b) Graph showing a comparison between the number
of devices collecting more than a certain charge Q (x-axis) in n-FinFETs
and p-FinFETs for 11MeV and 5MeV particle strikes.}
\end{figure}

We performed a similar experiment on an array of six bulk p-FinFETs
shown in Figure \ref{fig:p-finfets}. The number of devices affected
in the case of p-FinFETs is lesser than the number of devices affected
in a layout n-FinFETs. 

\subsection{Charge collection in a layout of multi-fin FinFETs}

\begin{figure}[!tbh]
$\;\;\;\;$\includegraphics[scale=0.5]{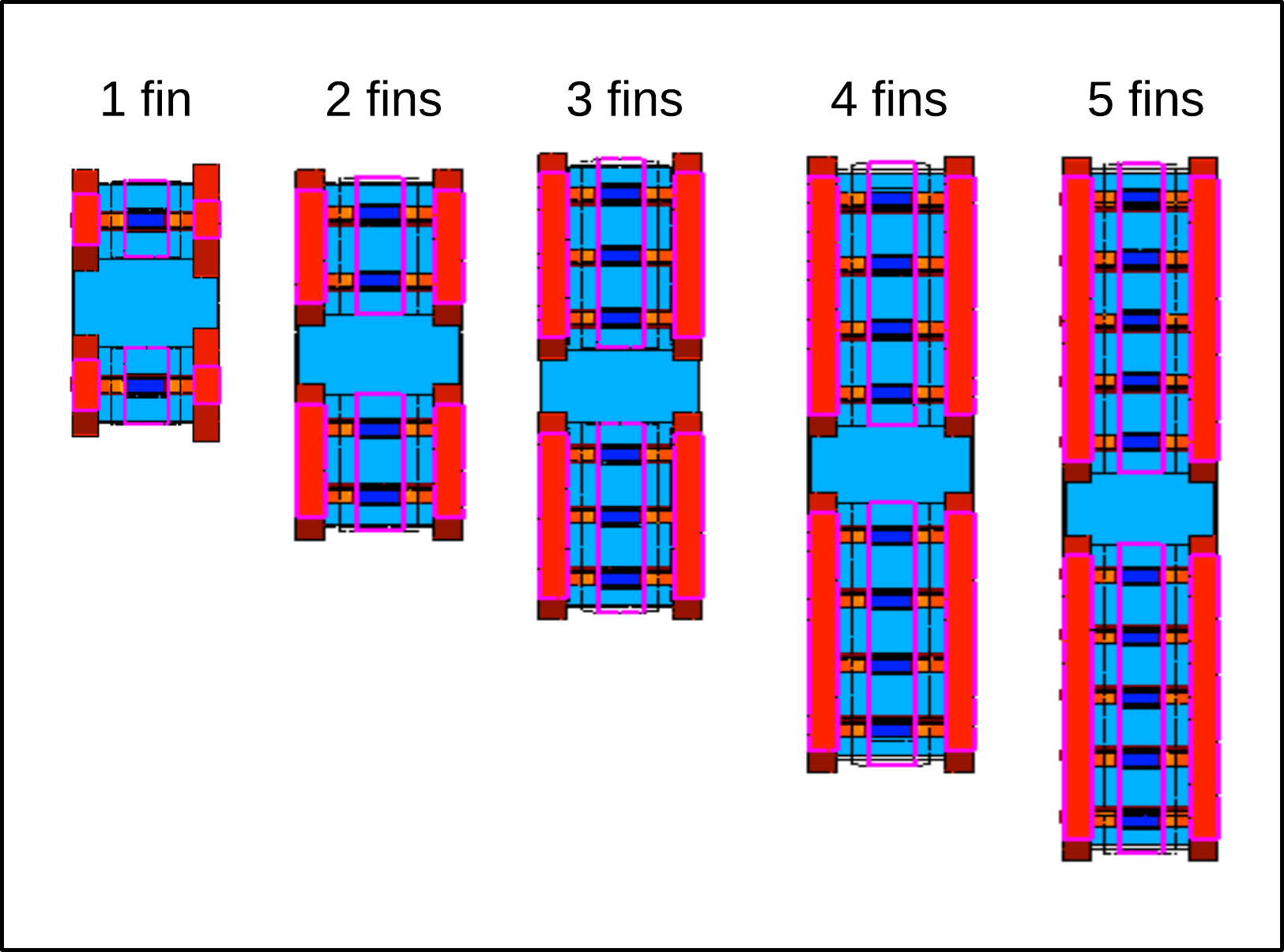}

\caption{\label{fig:multifin-finfets}Layout of two adjacent multi-fin bulk
n-FinFETs. The number of fins varies from one to five.}
\end{figure}

\begin{figure}[!tbh]
$\;\;\;\;\;\;$\includegraphics[scale=0.55]{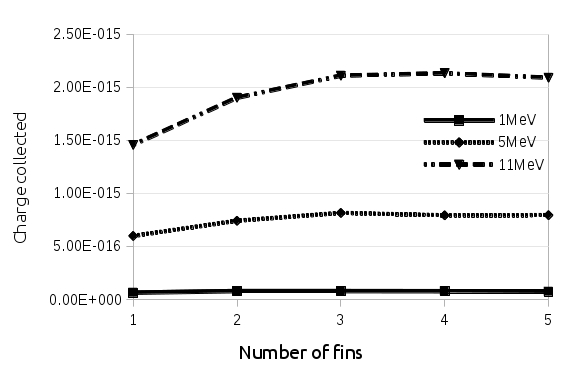}

\caption{\label{fig:multifin-finfets-charge}Charge collection in the FinFET
is plotted as the number of fins is varied}
\end{figure}

\begin{figure}[!tbh]
$\;\;\;\;\;\;\;\;\;\;\;$\includegraphics[scale=0.45]{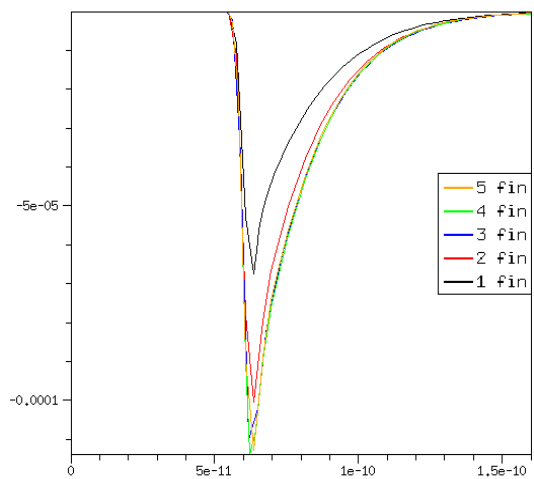}

\caption{\label{fig:multifin-finfets-transient}Transient current plots at
the drain of a FinFET as the number of fins is varied}
\end{figure}

Typically FinFETs have multiple fins. In this section, we study the
charge collection in a layout of such multi-fin FinFETs as the number
of fins is varied. So, we construct five different layouts of two
adjacent multi-fin FinFETs as shown in Figure \ref{fig:multifin-finfets}.
The number of fins in these five layouts varies from one to five respectively
as shown in the figure. We assume a 11MeV particle to be incident
in between the two devices and measure the charge collected in all
the source/drain regions in each case. In Figure \ref{fig:multifin-finfets-charge},
we plot the minimum of the collected charge in these source/drain
regions as the number of fins is increased. We see that a 2 fin FinFET
does not collect twice as much charge as a single fin FinFET; it collects
less. Similarly, a 5 fin FinFET does not collect five times as much
charge as a single fin FinFET. This can also be observed from the
transient current plots in Figure \ref{fig:multifin-finfets-transient}.
If we were to calculate the charge collected per fin, it is highest
in the case of two adjacent single fin FinFETs as compared to multi-fin
FinFETs. So, single fin FinFETs are thus more susceptible to particle
strikes.

\section{\label{sec:Conclusion}Conclusion}

We performed the analysis on 14nm bulk FinFETs to understand the extent
to which multiple transistors are affected in the current technology.
Our observations in the case of FinFETs are similar to those in the
case of planar transistors. Multiple transistors are affected due
to a single particle strike and the impact can last up to five transistors
away (up to 200nm). However, the nearest two transistors are the ones
that are most affected. Potentials of the source/drain regions have
a significant impact on charge collection: higher the potential, higher
is the charge collected. Further, we find that FinFETs with lesser
number of fins are more vulnerable to a particle strike. 

\section{Acknowledgements}

The authors would like to thank Sajesh Kumar for his help in introducing
us to the device tools and setup.

\bibliographystyle{ieeetr}
\phantomsection\addcontentsline{toc}{section}{\refname}\bibliography{Bibliography_nanditha}

\end{document}